\documentclass[aps,amssymb,prb,twocolumn,showpacs]{revtex4-1}
\usepackage{graphicx}
\usepackage{epsfig,amsmath}
\begin{document}
\def\eps{\varepsilon}
\def\TE{T_{\rm env}}
\title{Theory of single-electron heat engines coupled to electromagnetic environments}
\author{Tomi Ruokola$^1$}
\author{Teemu Ojanen$^{2,3}$}
\affiliation{$^1$ Department of Applied
Physics,  Aalto University, P.O.~Box 11100,
FI-00076 Aalto, Finland}
\affiliation{$^2$ Low Temperature Laboratory, Aalto University, P.O. Box 15100,
FI-00076 Aalto, Finland }
\affiliation{$^3$ Physics Department, Harvard University, Cambridge, Massachusetts 02138, USA}
\date{\today}
\begin{abstract}
We introduce a new class of mesoscopic heat engines consisting of a tunnel junction coupled
to a linear thermal bath. Work is produced by transporting electrons up against a voltage
bias like in ordinary thermoelectrics but heat is transferred by microwave photons,
allowing the heat bath to be widely separated from the electron system.
A simple and generic formalism
capable of treating a variety of different types of junctions and environments is presented.
We identify  the systems and conditions required for  maximal efficiency and maximal power.
High efficiencies are possible with quantum dot arrays but high power can be achieved also with
metallic systems.

\end{abstract}
 \bigskip
\maketitle

\section{Introduction}

Conversion of heat to work, and, in particular, recovery of waste heat
produced by electronic components is a problem of great and ever increasing
importance. Solid-state thermoelectric systems\cite{ali} are ideally suited for this purpose
since they are easily integrated with the rest of the circuitry
on the microchip. Mesoscopic heat transfer devices are a promising class
of thermoelectric systems due to their ease fabrication, control, and measurement,\cite{rmp}
and because sharp features in the energy spectrum, a requirement for efficient operation,\cite{mahan}
are readily available. Due to their small size they can also be used to study foundational
issues, such as the importance of fluctuations\cite{hanggi} and the fundamental limits of
heat engine performance.\cite{qubit}

Arguably the simplest mesoscopic heat engine consists of a single-level quantum dot placed between
two metallic leads held at different temperatures and voltages.\cite{linke}
Positioning the dot level far enough from the Fermi levels of the leads enables the electrons
to flow against the voltage bias while carrying heat from hot to cold. 
Thermoelectric properties of weakly coupled quantum dots have also been experimentally
studied.\cite{exp93,exp97,exp07,exp11}
This type of device operates as a heat engine by generating electrical current
and transferring heat between the same two reservoirs.
A recent modification\cite{rafa1,rafa2} of this scheme makes the charge and heat currents
flow along different pathways
by introducing a third reservoir: charge is transported between two reservoirs at the same temperature,
while the third reservoir, at a different temperature, is Coulomb-coupled to the transport
electrons and supplies the thermal fluctuations driving the heat engine.

\begin{figure}[t]
\centering
\includegraphics[width=.75\columnwidth,clip]{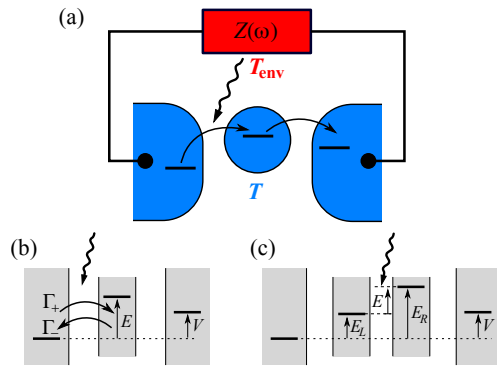}
\caption{(color online). (a) Schematic of a single-electron
heat engine coupled to an external environment with impedance $Z(\omega)$. Electrons tunneling
through the left junction exchange energy with the heat bath, enabling
net current against the voltage bias. (b), (c) Energy-level diagrams of junction systems with one
or two dots between the leads. One of the junctions, marked with a photon symbol,
is coupled to the external bath. The photon-assisted tunneling rates are $\Gamma_\pm$.
}\label{fig1}
\end{figure}

Here we introduce a new type of mesoscopic heat engine, sketched in Fig.~\ref{fig1}(a).
It consists of one or more quantum dots or metallic dots between two electronic leads at
the same temperature $T$ but with a voltage bias $V$.
The leads are connected to an external circuit with temperature
$\TE$ and impedance $Z(\omega)$. If the tunnel coupling for one of the junctions is small
enough, a tunneling electron will exchange energy with the electromagnetic environment, and with a proper
choice of parameters it is then possible to achieve a net current against the voltage bias.
This system is similar in spirit to the devices of Refs.~\onlinecite{rafa1,rafa2} but instead
of a direct Coulomb interaction between the electrons in the transport system and in the
heat bath, in our proposal the coupling is mediated by microwave photons.
Photonic heat conduction in electronic circuits has been investigated in recent years
both theoretically\cite{schmidt,niscooler,teemu1,teemu2,teemu3,hekking}
and experimentally,\cite{jukka1,jukka2}
while previously studied single-electron devices for thermal applications
include a thermometer\cite{rmp,cbt}, a cooler\cite{niscooler},
and a heat diode.\cite{diode}
There are also theoretical proposals for three-reservoir heat engines where the electrons
are driven by a coupling to a phonon bath.\cite{ora,peng}

A significant
advantage of the photonic coupling compared to Coulombic or phononic interaction
is that the two parts of the circuit with different
temperatures can be located
arbitrarily far apart. One can imagine a scenario where the external circuit is a relatively large
device which performs some useful function but at the same time produces excess heat.
Our engine can recover a part of this heat and feed it back to the main device as
electrical power.

We show that in an optimal configuration the electrical current in different types of junction
systems is given  by a single concise formula, Eq.~(\ref{currsimp}).
Then a junction between two quantum dots is shown to be ideal in terms of efficiency
while all junction types, involving either metals or quantum dots, are able to deliver
approximately equal maximal power production.

The rest of the paper is organized as follows. In Sec.~\ref{petheory} we  present the
theory for electron tunneling in a linear external environment.
In Sec.~\ref{junctions} we consider bath-assisted transport in
junction arrays and conclude that in an appropriate limit the current
is given by Eq.~(\ref{currsimp}) for a variety of different
systems. The heat engine characteristics of different junction types are studied
in detail in Sec.~\ref{engine}, with a particular emphasis on  performance
under maximum power conditions.
In Sec.~\ref{conc} we summarize the results and consider experimental prospects.

\section{Coupling to the environment}\label{petheory}

The exchange of energy between the tunneling electron and the external
environment is treated with the so-called $P(E)$ theory. A thorough
account of this formalism is given in Refs.~\onlinecite{naz1,naz2},
and in this Section we only present the results relevant for the present
study.
$P(E)$ theory, and therefore also our present work, rests on two
fundamental requirements: i) it is assumed that the coupling between
the different electron systems is weak enough and that the temperature
of the electrons or the environment is high
enough so that transport can be adequately described by the lowest-order
Fermi golden rule, and ii) it is assumed that the environment relaxation
is much faster than the tunneling rate.
Then within $P(E)$ theory
the tunneling rate $\Gamma_{i\to j}$ between electron systems
$i$ and $j$ through a junction, that is, a single insulating barrier,
is given by the Fermi
golden rule formula with the energy-conserving $\delta$ function
replaced by $P(E)$, the probability density to exchange energy $E$
with the environment. Thus we have\cite{naz1}
(with $\hbar=k=e=1$)
\begin{equation}\label{gamma}
\Gamma_{i\to j}=2\pi |t|^2\int d\eps_id\eps_j
\rho_i(\eps_i-\mu_i)\bar{\rho}_j(\eps_j-\mu_j)P(\eps_{ij})
\end{equation}
where $\eps_{ij}=\eps_i-\eps_j$.
For the electron density $\rho_i(\eps)$ we consider two cases,
a single-level quantum dot with $\rho_i(\eps) = \delta(\eps)$, and a metal
with $\rho_i(\eps) = \nu_if(\eps)$ where $\nu_i$ is the density of states
and $f(\eps)$ is the Fermi function. Similarly the hole density is
$\bar{\rho}_i(\eps) = \delta(\eps)$ for dots and $\bar{\rho}_i(\eps) = \nu_i[1-f(\eps)]$
for metals. Note that in the quantum dot case we assume that dot $i$ is occupied
and dot $j$ is empty, otherwise the rate would vanish.
The tunneling matrix element between the initial and final states
is $t$, which is taken here to be energy independent.
The Fermi level of a metal or the single
level of a quantum dot is $\mu_i$ with possible Coulomb charging energies
absorbed into it. The heat current emitted by the environment during the tunneling process
is obtained from the rate formula by weighing the integral with
$\eps_j-\eps_i=-\eps_{ij}$:
\begin{equation}\label{jij}
J_{i\to j}=-2\pi |t|^2\int d\eps_id\eps_j\,\eps_{ij}
\rho_i(\eps_i-\mu_i)\bar{\rho}_j(\eps_j-\mu_j)P(\eps_{ij})
\end{equation}

The $P(E)$ function for an electromagnetic environment
can be determined by a circuit theory analysis
of the system.
If the junction, which itself has some capacitance $C$,
is coupled to an environment with
impedance $Z(\omega)$,
the total impedance $Z_t$ over the junction is $C$ and $Z(\omega)$ in parallel,
that is, $Z_t(\omega)=[i\omega C+1/Z(\omega)]^{-1}$.
We will omit the rather complicated general expression for $P(E)$.
For our purposes it is sufficient to note that it only depends on the
environment temperature $\TE$ and the real part of $Z_t(\omega)$.
Since $P(E)$ is a probability density its integral is normalized to unity,
and additionally the detailed balance for the
environment requires that\cite{naz1} $P(-E)=e^{-E/\TE}P(E)$.

When presenting numerical results we will consider a
simple and prototypical environment, namely an ohmic resistor
with impedance $Z(\omega)=R$. We further assume the high-impedance
limit where $RC$ is the largest time scale of the system.
Then the $P(E)$ function is given as\cite{naz1}
\begin{align}\label{highz}
P(E) = \frac{1}{\sqrt{2\pi\sigma^2}}e^{-\frac{(E-E_C)^2}{2\sigma^2}}
\end{align}
where $E_C=e^2/2C$ is the charging energy of a junction
with capacitance $C$ and $\sigma^2=2E_C\TE$.

\section{Currents in junction systems}\label{junctions}

Thermoelectric power generation in a three-reservoir single-electron
device requires at least one environment-coupled tunnel junction,
and we will now consider systems with one, two, or three junctions
in series. The concrete realizations of these systems are
zero, one, or two (quantum or metallic) dots placed between two metallic reservoirs.
We work within the framework of $P(E)$ theory, and thus assume weak coupling
and fast environment relaxation, as explained in Sec.~\ref{petheory}.

Let us start with a single junction with no dots between
two noninteracting metallic leads at temperature $T$. The junction is coupled to an electromagnetic
environment with
an arbitrary $P(E)$ and temperature $\TE$. The tunneling rate for the positive direction,
from left ($L$) to right ($R$),
is $\Gamma_+$ and the rate in the opposite direction is $\Gamma_-$.
With the voltage bias $\mu_R-\mu_L=V$,
Eq.~(\ref{gamma}) gives the current as
\begin{align}\label{simple}
I_0&=\Gamma_+-\Gamma_-\\
&=\gamma\int d\eps_Ld\eps_R\,
f(\eps_L)[f(V-\eps_R)-f(-V-\eps_R)]P(\eps)\nonumber
\end{align}
where $\gamma=2\pi|t|^2\nu_L\nu_R$ and $\eps = \eps_L-\eps_R$.
The subscript $0$ for the current denotes the fact that there are no dots in the system.
We use a convention where a positive voltage $V$ means that electrons have a
higher energy in the right lead and a positive current $I$ means that electrons
travel, on the average, from left to right. Therefore if $I$ and $V$ have
the same sign, that is, the generated power $\dot{W}=IV$ is positive, heat is converted
to electrical work and the device operates as a heat engine.
For the present case the bracketed term in Eq.~(\ref{simple}) has a sign that is opposite to the sign
of $V$ and therefore $\dot{W}\le0$. This simple junction
cannot produce thermoelectric power. One way to understand this
failure is to think of the system as a Brownian motor\cite{reimann}
which is driven by
thermal fluctuations due to the external environment. However, these
fluctuations do not intrinsically have any
preferred direction and therefore
the noise must be rectified by the electron transport system
in order to generate net power.\cite{rafa2}
Rectification requires a nonlinear
current--voltage characteristic but the simple junction (without the
environment coupling)
is linear. Therefore a nonlinearity must be introduced into
the transport system, and here we do it by
adding a (quantum or metallic) dot between the two
leads. We further assume that the Coulomb repulsion within the dot
is so strong that it can only be empty (with probability $p_0$)
or singly occupied (with probability $p_1$).
The Fermi level of the metallic dot or the single-particle level of 
the quantum dot is at energy $E$, with Coulomb energy included;
see Fig.~\ref{fig1}(b).
Now the system has two junctions, and the tunneling rates
through the left and right junction are $\Gamma_\pm$ and
$\Gamma_{R\pm}$, respectively.
Instead of Eq.~(\ref{simple}), the current is given by
$I=p_0\Gamma_+-p_1\Gamma_-$. The probabilities
can be solved from the master equation
$\dot{p}_0=p_1(\Gamma_-+\Gamma_{R+})-p_0(\Gamma_++\Gamma_{R-})$
in the steady state, $\dot{p}_0=0$.
This gives the current for the one-dot system as
\begin{align}\label{curr2}
I_1=\frac{\Gamma_+\Gamma_{R+}-\Gamma_-\Gamma_{R-}}{
\Gamma_++\Gamma_{R+}+\Gamma_-+\Gamma_{R-}}
\end{align}
If the two junctions are identical, symmetry of the system implies
that there is no thermally induced net current for $V=0$, and a 
calculation with Eq.~(\ref{gamma}) shows that there can be no power
production for any $V$. For instance, in the case of a quantum dot
the numerator of Eq.~(\ref{curr2}) is proportional to
$\int d\eps d\eps^\prime\,P(\eps)P(\eps^\prime)
[f(\eps+E)f(\eps^\prime-E+V)-f(\eps+E-V)f(\eps^\prime-E)]$.
For $V>0$ the first and the second Fermi function are smaller that the
third and the fourth, respectively, implying $I_1<0$.
To arrive at the preceeding expression the $P(E)$ functions for the two junctions
must be identical, and therefore unequal environmental couplings are required
for power production.

Since both junctions are connected to the same external circuit,
the only way they can have different $P(E)$ functions is that the
junction capacitances, $C_L$ and $C_R$, are unequal.
In that case circuit theory analysis\cite{naz1}
shows that the total impedance over
junction $i$ is $\textrm{Re}\,Z_{t,i}(\omega)=\kappa_i^2\textrm{Re}\,\tilde{Z}_{t}(\omega)$,
where $\kappa_i=C/C_i$ and $\tilde{Z}_{t}(\omega) = [i\omega C+1/Z(\omega)]^{-1}$,
with $1/C=1/C_L+1/C_R$.
In other words, the only difference to the single junction case
is that the external impedance seen by junction $i$ is effectively
reduced by a factor $\kappa_i^2$.
Thus a maximal difference between the $P(E)$ functions of the two junctions can be obtained
when the capacitances have different orders of magnitude.
If, for example, with a small left-side capacitance $C_L\ll C_R$, we have $C\approx C_L$, $\kappa_L\approx1$,
and $\kappa_R\approx C_L/C_R$. In this limit the left junction is coupled
to the environment with energy $E_C=e^2/2C_L$ while the coupling of the right junction
is suppressed by a factor of $(C_L/C_R)^2$. Note however that the charging energy
of the dot is $e^2/2(C_L+C_R)$ which must still be large enough to prevent a multiple occupation of the dot.
To simplify Eq.~(\ref{curr2}) we note that since now
the right junction is effectively decoupled from the environment, detailed balance requires
$\Gamma_{R+}=\Gamma_{R-}e^{(E-V)/T}$. This relation can, of course, be also derived
from Eq.~(\ref{gamma}) with $P(E)=\delta(E)$.
Furthermore, a junction with a small capacitance is usually also weakly transmitting and therefore it
is consistent to assume that $C_L\ll C_R$ implies $\Gamma_\pm\ll\Gamma_{R\pm}$.
Then we obtain
\begin{equation}\label{currsimp}
I=f(V-E)\Gamma_+ - f(E-V)\Gamma_-
\end{equation}
This is the main equation of the present paper. Below we argue that it applies
also to more generic junction systems and therefore we have dropped the
subscript from $I_1$. The heat engine behavior produced by this equation is examined
in detail in the next section.

Let us then turn to the case of two dots in series between
the leads, as depicted in Fig.~\ref{fig1}(c).
The energy levels of the left and right dot are $E_L$ and $E_R$,
respectively, and $E=E_R-E_L$.
We again assume a strong enough Coulomb repulsion so that the double dot
can be either empty (probability $p_0$), or there can be one electron in the left
dot (probability $p_L$) or in the right dot (probability $p_R$). The tunneling rates
through the left, center, and right junction are $\Gamma_{L\pm}$, $\Gamma_{\pm}$,
and  $\Gamma_{R\pm}$, respectively, and the current is given by
$I_2=p_L\Gamma_+-p_R\Gamma_-$. Master equation for the occupations is
\begin{eqnarray}
\dot{p}_0&=&-p_0(\Gamma_{L+}+\Gamma_{R-})+p_L\Gamma_{L-}+p_R\Gamma_{R+}\nonumber\\
\dot{p}_L&=&p_0\Gamma_{L+}-p_L(\Gamma_{L-}+\Gamma_{+})+p_R\Gamma_-\\
\dot{p}_R&=&p_0\Gamma_{R-}+p_L\Gamma_{+}-p_R(\Gamma_{R+}+\Gamma_-)\nonumber
\end{eqnarray}
and the steady-state solution gives the current as
\begin{equation}
I_2 = (\Gamma_{L+}\Gamma_{+}\Gamma_{R+}-\Gamma_{L-}\Gamma_{-}\Gamma_{R-})/\tilde{\Gamma}^2
\end{equation}
with $\tilde{\Gamma}^2=\Gamma_{L+}\Gamma_{R+}+\Gamma_{L-}\Gamma_{R-}+
\Gamma_{L-}\Gamma_{R+}+\Gamma_{+}(\Gamma_{L+}+\Gamma_{R+}+\Gamma_{R-})+
\Gamma_{-}(\Gamma_{L+}+\Gamma_{L-}+\Gamma_{R-})$. Similarly to the one-dot
case, we now assume that the central junction between the dots has a small capacitance and
a small transmittance compared to the other two junctions. Thus $\Gamma_{\pm}\ll
\Gamma_{L\pm,R\pm}$, and since now the left and right junctions are effectively decoupled
from the environment, detailed balance implies $\Gamma_{L-}/\Gamma_{L+}=e^{E_L/T}$
and $\Gamma_{R+}/\Gamma_{R-}=e^{(E_R-V)/T}$. The expression for the current is
then simplified to
\begin{equation}\label{curr3b}
I_2 = \frac{e^{(E-V)/T}\Gamma_+-\Gamma_-}{1+e^{(E-V)/T}+e^{(E_R-V)/T}}
\end{equation}
The last term of the denominator is vanishingly small compared to the first
two if $-(E_R-V)\gg T$ and $-E_L\gg T$. In this limit the expression for
the current is again reduced to Eq.~(\ref{currsimp}). Physically this limit
means that the dot levels are much below the Fermi levels of the leads
and thus either of the two dots is always occupied ($p_0\to0$), leading effectively
to a two-state system similar to the one-dot case. Clearly this argument could
also be extended to a larger number of dots if necessary.

We have now seen that  Eq.~(\ref{currsimp}) is the fundamental expression
for thermoelectric current in the optimal limit when only one junction is
exchanging heat with the external bath and when all ``idle'' time spent on processes
with no environment coupling can be neglected.
The generality of Eq.~(\ref{currsimp}) can also be intuitively seen as follows.
The system is always in one of two possible states:
there is an electron ready to tunnel either from left to right or from right
to left through the bath-coupled junction.
The probabilities of these
two states are $p_+$ and $p_-$, and the average current through the junction
is therefore $I=p_+\Gamma_+ - p_-\Gamma_-$. On the left side of the junction
there is a metal with Fermi level $E_L$, or a quantum dot with a level at $E_L$
strongly coupled to a metal. In either case, the probability that the left side
has an electron ready for tunneling is proportional to $f(E_L)$. Similarly
the right side has a level at $E_R-V$, and it is ready to receive the tunneling
electron with a probability proportional to $1-f(E_R-V)=f(V-E_R)$. Thus we
have $p_+\propto f(E_L)f(V-E_R)$ and analogously
$p_-\propto f(-E_L)f(E_R-V)$. Normalizing with
$p_++p_-=1$ gives $p_+=f(V-E)$ and $p_-=f(E-V)$, and we arrive at Eq.~(\ref{currsimp}).
The existence of these two states is due to the electron--electron interaction.
The noninteracting case of Eq.~(\ref{simple}) is obtained when
there is only one single state allowing electrons to tunnel at any time in either
direction.

The heat current $J$ emitted by the environment can be calculated by following
exactly the same steps as for the electrical current, and the result
corresponding to Eq.~(\ref{currsimp}) is
\begin{equation}\label{jsimp}
J=f(V-E)J_+ + f(E-V)J_-
\end{equation}
where $J_+$ and $J_-$ are the energy currents absorbed by the electron tunneling
to the right and to the left, respectively,
as obtained from Eq.~(\ref{jij}). The sign difference between Eqs.~(\ref{currsimp})
and (\ref{jsimp}) is due to the fact that $J_\pm$ already contain the direction of the heat flow.

\section{Heat engine characteristics}\label{engine}

As we have argued, in the optimal limit an array of one or more dots
can be seen as a two-state system obeying Eqs.~(\ref{currsimp})
and (\ref{jsimp}), and thus when evaluating the thermoelectric power generation
in these devices it is only necessary to consider the
single junction that is coupled to the electromagnetic environment.
Some generic remarks about Eq.~(\ref{currsimp}) can be made without any explicit
model for the junction. First note that a mirror reflection of a solution with $(E,V,I)$
produces another solution with $(-E,-V,-I)$ and therefore it is sufficient
to consider the case $E>0$. Next, the $I(V)$ curve of Eq.~(\ref{currsimp})
has a simple overall structure.
Since the tunneling rates $\Gamma_\pm$ depend only on the level difference $E$
over the junction but not on the voltage bias $V$, the current
depends
on $V$ only through the Fermi functions. Thus we see that $I$ approaches asymptotically
$\Gamma_+$ and $-\Gamma_-$ for large negative and positive values of $V$, respectively.
Since $I(V)$ decreases monotonically, at some point $V=V_0$ the current vanishes.
For our purposes the most important fact is that
for voltages between $0$ and $V_0$, $I$ and $V$ have the same sign and the device operates
as a heat engine. From Eq.~(\ref{currsimp}) we can solve
\begin{equation}\label{v0}
V_0=E-T\log\frac{\Gamma_-}{\Gamma_+}
\end{equation}
All the systems studied here have $\Gamma_->\Gamma_+$ and therefore $E$
is an upper limit for $V_0$ while no lower limit exists. We will also see
that a hot environment ($\TE>T$) implies $V_0>0$ and therefore the heat engine operates with
positive $V$ and $I$ while for a cold environment ($\TE<T$) $V$ and $I$ are negative.

An important characteristic of a heat engine is the efficiency $\eta=IV/J_H$, where $J_H$ is the heat current
from the hot bath. When $\TE>T$ we have simply $J_H=J$ but when $\TE<T$ the hot
bath is the transport system and then $J_H=IV-J$, that is, the heat taken from
the electrons is the sum of the produced power $IV$ and the heat $-J$ expelled
to the cold environment. Thus we have
\begin{equation}\label{eff}
\eta = \left\{
\begin{array}{ll}
\frac{IV}{J}, & \TE>T\\
\frac{IV}{IV-J}, & \TE<T
\end{array}\right.
\end{equation}
The fundamental upper limit for $\eta$ 
is the Carnot efficiency $\eta_C=1-T_C/T_H$, where $T_C=\min\{T,\TE\}$ 
and $T_H=\max\{T,\TE\}$. In this limit transport proceeds reversibly and
power production is vanishingly small. Thus reaching Carnot efficiency is not
a very useful goal in practice.

Another efficiency measure, widely used in the context of
thermoelectric power generation, is the figure
of merit $ZT$. Even though the separation of heat and charge pathways makes the devices
covered by our theory quite different from typical thermoelectrics, the usual definition
of $ZT$ can still be straightforwardly applied to the present case.
In the linear response limit, when $V$ and $\Delta T\equiv \TE-T$ are much smaller
than $T$, we have\cite{ali}
\begin{equation}\label{zt}
ZT=\frac{\sigma S^2}{\kappa}T
\end{equation}
where the Seebeck coefficient is $S=\partial V_0/\partial(\Delta T)$,
 electrical conductance is $\sigma=\partial I/\partial V$ at $\Delta T=0$,
and thermal conductance is $\kappa=\partial J/\partial(\Delta T)$ at $I=0$.
Generally $\kappa$ should include all forms of heat transfer between
the reservoirs, but
since we are not modeling any parasitic flows they are not included in $ZT$.
There is a major effort in thermoelectrics research to produce systems with $ZT>1$.\cite{ali}

When an electron is transported between the leads of a thermoelectric system
it always performs the same amount of work $\pm V$ where the sign depends on the direction
of tunneling. However, the amount of heat transferred between the thermal baths
can vary from one tunneling event to another, and the spectrum of this energy
exchange is determined both by the structure of the junction and the environment.
We consider two limiting cases for the junction structure. The first is a fully
energy-selective junction, where each tunneling event is accompanied by an
exchange of a fixed amount of heat $\pm E$. A physical realization for this
is a junction between two quantum dots with sharply defined energy levels
separated by $E$. The other extreme is a totally unfiltered junction
where essentially any amount of heat can be exchanged. The two physical
examples that we consider are a junction between a quantum dot and a metal,
and a junction between two metals.

\subsection{Energy-selective junctions}

A junction between two quantum dots can only exchange a fixed energy $E$ with
the external bath during the tunneling events. Using Eq.~(\ref{gamma})
together with detailed balance for the
environment yields
\begin{equation}\label{gammaqd}
\begin{array}{lcl}
\Gamma_+&=&\Gamma_-e^{-E/\TE}\\
\Gamma_- &=& 2\pi|t|^2P(E)
\end{array}
\end{equation}
 Similarly
the heat flows from  Eq.~(\ref{jij}) are $J_+=E\Gamma_+$ and $J_-=-E\Gamma_-$,
which imply the simple relation $J=EI$. Therefore at all temperatures the
ratio of produced power and transferred heat is a constant $V/E$,
a situation known as strong coupling between particle and heat flows.
Carnot efficiency can only be achieved by such strongly-coupled systems,\cite{qubit,kedem,broeck}
and in the present case this can be confirmed by noting that
the stopping voltage from Eq.~(\ref{v0}) is
\begin{equation}\label{v0qd}
V_0=E(1-T/\TE)
\end{equation}
and substituting $V=V_0$ in Eq.~(\ref{eff}) yields $\eta=\eta_C$.
Thus Carnot efficiency implies vanishing current and power.
The value for the figure of merit $ZT$ can be inferred directly by noting that 
since $\kappa$ in Eq.~(\ref{zt}) is evaluated at $I=0$, the strong-coupling
condition implies $J=\kappa=0$ and therefore $ZT=\infty$, independent of
any parameters. This is another indication that energy-selective junctions
can be used to construct maximally efficient thermoelectrics.

In order to investigate the physics beyond linear response and to find
out the conditions that maximize the power instead of efficiency,
we perform a numerical calculation and for that purpose an explicit expression
for the environment spectrum is needed.
Since an energy-selective junction interacts with the environment
only at a single energy $E$, the full form of the $P(E)$ function does not generally have
much significance. There is one caveat, however: Fermi golden rule, which $P(E)$ theory
is based on, is not able to treat transitions between discrete states, and therefore
our approach fails if the environment and both sides of the junction are discrete.
Our example environment of Eq.~(\ref{highz}) is continuous and therefore the present
approach is valid also for a junction between two quantum dots, except in the limit $\sigma\to0$
when $P(E)$ becomes a discrete delta peak.

\begin{figure}[t]
\centering
\includegraphics[width=\columnwidth,clip]{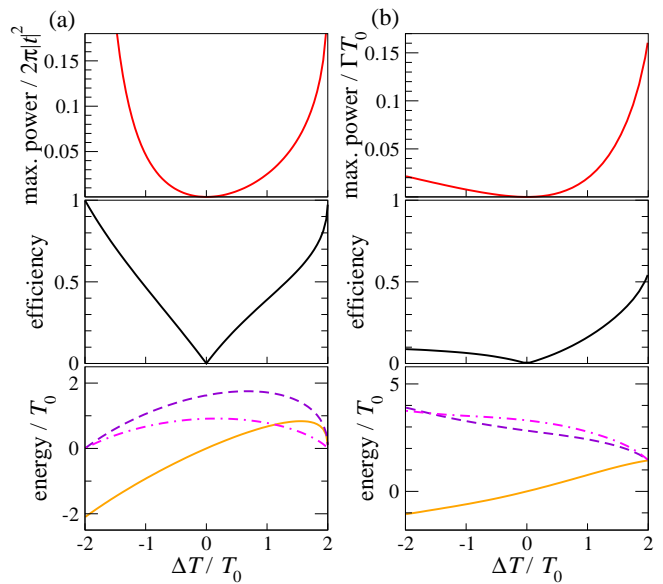}
\caption{(color online). Maximum power production for
(a) a junction between two quantum dots, and
(b) a junction between a quantum dot and a metal.
Top panels: The maximum power for given temperatures. Note the different
units of power for (a) and (b). Middle panels: efficiency at maximum power.
Bottom panels: values of $V$ (full line), $E$ (dashed), and $E_C$ (dash-dotted) that give the maximum
power. The temperature difference is $\Delta T=\TE-T$ and the average temperature
is $T_0=\frac 1 2(T+\TE)$.
}\label{fig2}
\end{figure}

For each pair of temperatures $T$ and $\TE$, the generated power $\dot{W}=IV$
from Eqs.~(\ref{currsimp}) and (\ref{gammaqd}) is numerically maximized
with respect to the bias $V$, the level difference $E$,
and the coupling energy $E_C$, and the results are presented
in Fig.~\ref{fig2}(a).
In order to estimate the maximum power achievable
with this device we note that
the unit of power in Fig.~\ref{fig2}(a) is $2\pi|t|^2$, and to find an upper limit for this quantity
we model the junction between two quantum dots as a two-state
system with level difference $E$ and tunnel coupling $t$. The left and right dots must be approximate
energy eigenstates, requiring that $|t|\ll E$. Since $E\approx T_0$ for maximum power operation,
we end up with $|t|\ll T_0$. Thus the output power is strictly limited by the operating temperature.
The spurious divergence of power for very large temperature differences is due to the
breakdown of $P(E)$ theory, as explained in the preceeding paragraph.

We remark that this energy-selective system is very similar to the
one studied in Ref.~\onlinecite{rafa1}. The physical implementations
are rather different, with the system of Ref.~\onlinecite{rafa1} requiring a total
of four quantum dots  while for our device
two dots are sufficient, but the structures of the energy transfer processes are fundamentally identical,
leading to very similar performance figures.

\subsection{Unfiltered junctions}

Junctions which do not restrict the amount of energy
exchanged between the tunneling electron and the environment will be called 
\emph{unfiltered}. We concentrate on a system
where the other side of the junction is a quantum dot and the other side
is a metal, but at the end of the Section we also briefly consider a junction between
two metals. For a metal--quantum dot junction  Eq.~(\ref{gamma}) gives
\begin{equation}\label{gammam}
\Gamma_\pm=\Gamma\int d\eps P(\eps)f(\eps\pm E)
\end{equation}
where $\Gamma=2\pi|t|^2\nu$. This form shows immediately that $\Gamma_->\Gamma_+$
and thus $V_0<E$. 
Linearizing Eqs.~(\ref{currsimp}) and (\ref{jsimp}) then yields
the figure of merit from Eq.~(\ref{zt}) as
\begin{equation}
ZT=\left(\frac{c\tilde{J}}{\tilde{E}^2}-1\right)^{-1}
\end{equation}
where $c=F[0]$, $\tilde{E}=F[1]$, $\tilde{J}=F[2]$,
and $F[n] = \int_0^\infty d\eps P(\eps)\eps^n[f(E-\eps)e^{-\frac{\eps}{T}} +(-1)^n f(E+\eps)]$.
These definitions show that if $P(E)$ consists of a pair of delta peaks at
energies $\pm E_0$, then in the limit $E=E_0\gg T$ we have $ZT\to\infty$
and therefore this kind of environment is able to mimic the effect of an energy-selective
junction. However, in this limit $\tilde{E}$, and thus the power $IV$, scales as
$\exp(-E/T)$. A more detailed analysis shows that beyond $ZT\sim1$ a linear increase
in $ZT$ corresponds to an exponential suppression of generated power. This should be contrasted
to the energy-selective junction where the infinite $ZT$ is due to the idealized assumption
of perfectly sharp energy levels. In reality the levels are broadened and the figure of merit
is finite. However, when the level width is decreased, power is reduced roughly inversely, and not
exponentially, with $ZT$.\cite{linke_max} Thus an energy-selective junction is the only
realistic way of achieving very high efficiencies, at least in the linear regime.
The fundamental difference between the energy filtering provided by the junction
and by the environment is the fact the former is directional while the latter is not:
when an electron tunnels to the right in an energy-selective junction it must always absorb 
a photon and when tunneling in the other direction it must emit a photon, but
for an unfiltered junction both emission and absorption are
possible for either direction.

After concluding  that large efficiencies are not available for linear response,
we turn to the nonlinear regime and conditions for maximum power.
Figure~\ref{fig2}(b) shows numerical results for an ohmic environment, as represented
by Eq.~ (\ref{highz}), with the power calculated from Eqs.~(\ref{currsimp}) and (\ref{gammam}).
The figure of merit at maximum power
is $ZT\approx 0.5$.
For $\TE>T$ the power is large but the efficiency clearly below the energy-selective case.
This is because the power is generated by elementary tunneling processes each performing
the same work $V$ but absorbing a different amount of heat from the hot bath.
These elementary processes thus have different efficiencies, and the overall average
efficiency will be lower than in the energy-selective case where each process
carries the same amount of heat.

In the opposite case of a cold environment, $\TE<T$, performance of the heat engine
is dramatically degraded.
Since now $V<0$, electrons tunneling from right to left perform
useful work, and therefore large power production requires that $\Gamma_-$ dominates $\Gamma_+$.
Indeed this is the case for the energy-selective junction: as can be seen from
Eq.~(\ref{gammaqd}), $\TE\to0$ implies $\Gamma_-\gg\Gamma_+$.
However, the unfiltered case of Eq.~(\ref{gammam}) does not share this property.
The physical explanation is that in the former case electrons can tunnel from left to
right only by absorbing energy from the cold environment while in the latter
case they can tunnel by using the thermal energy
of the hot electron system, without any energy exchange with  the external bath.
The transport processes which are decoupled from the environment produce
a considerable leakage current down along the voltage bias, thus making heat engine
performance very poor.

For a metal--quantum dot junction,
the magnitude of the maximum power depends on the product of $\Gamma$ and  $T_0$.
Validity of our approach requires that sequential tunneling dominates all higher-order
processes, and this is the case if $\Gamma\ll T_0$, and thus the unit of
power is constrained by $\Gamma T_0\ll T_0^2$.

Another model system for the unfiltered case
is a tunnel junction between two metals. In this case Eq.~(\ref{gamma}) yields
\begin{equation}\label{gammamm}
\Gamma_\pm=\gamma\int d\eps P(\eps)(\eps\pm E) n(\eps\pm E)
\end{equation}
where $\gamma=2\pi|t|^2\nu_L\nu_R$ and $n(\eps)$ is the Bose function
at temperature $T$.  We have also used the identity
$\int d\eps^\prime f(\eps^\prime)[1-f(\eps^\prime-\eps)]=\eps n(\eps)$.
Comparing Eq.~(\ref{gammamm}) to Eq.~(\ref{gammam}) we see that
the quantitative results for the metal--quantum dot junction can
be transferred to the present case by replacing $\Gamma\to\gamma$
and $f(\eps)\to\eps n(\eps)$. All qualitative arguments remain unchanged,
and numerical maximization yields results very similar to those in
Fig.~\ref{fig2}(b), with the unit of power now  being $\gamma T_0^2$.
The existence of Coulomb blockade requires\cite{naz2} $\gamma\ll 1$ and
therefore just like in the previous cases
the operating temperature strictly limits the attainable
power.

\section{Discussion}\label{conc}

Above we have shown that
highly efficient energy conversion is in practice only available
for an energy-selective junction. This type of device also performs equally
well for hot and cold environments. On the other hand, unfiltered junctions should
be operated with a hot environment and cold electron transport system.
To intuitively understand this behavior,
one can consider the system in Fig.~\ref{fig1}(b) with the left junction coupled to
the environment. When the electron system is cold, that is,
$T$ is small compared to the other energies,
then the Fermi functions
are sharp and electrons cannot tunnel up in energy from the right lead to the center,
and therefore only the positive direction transport processes remain. First an electron tunnels
from the left lead to the center by absorbing a photon, and then due to the strong
tunnel-coupling of the second junction it discharges to the right lead. Thus each
tunneling electron transfers heat from hot to cold and performs useful work,
leading to optimal thermoelectric performance.
On the other hand, for a hot electron system the Fermi functions are smeared and
in general electrons
can tunnel in both directions without absorbing or emitting photons. Such environment-decoupled
processes transport electrical current down the voltage bias, that is, they produce Joule heating from work,
hence severely degrading the engine power and efficiency.
Only a fully  energy-selective junction is able to force the
tunneling electrons to always interact with the environment, resulting in highly
efficient energy conversion.

Achieving high efficiency is typically a major goal in heat engine design
but for the purposes of waste heat recovery the thermal input energy of the device can be considered
free and abundant, making efficiency an irrelevant quantity.
In this case one should instead concentrate on maximizing the power output,
as has been done in Sec.~\ref{engine}.
From Fig.~\ref{fig2} one can see that the attainable power is roughly
$0.01\dots0.1$, expressed in units of 
$2\pi|t|^2$ for a junction between quantum dots, $\Gamma T_0$ for a metal--quantum dot junction,
and $\gamma T_0^2$ for a metal--metal junction. We have concluded that these three expressions
are all bounded to be much smaller than $T_0^2$, which leads us to estimate
that the maximum power achievable with this type of device is about $10^{-2}T_0^2$.
If $T_0=1\ K$, the generated power
falls in the femtowatt range.
This is a typical figure for low-temperature single-electron devices.\cite{diode,linke_max}

In principle it is possible to increase the power production by having several
engine units in parallel with the external impedance. The coupling energy $E_C$
scales inversely with the number of parallel devices and therefore
the capacitances of the individual junctions should be decreased. However, in order to
see an actual increase in current and power, this change in the capacitances should
not considerably lower the tunneling rates of the junctions.

Even if one is not concerned with efficiency, heat leaks between the thermal baths
should be minimized in order to have a maximally large temperature difference.
Phononic heat conduction is a problem with all thermoelectric systems. Since phonons
are not part of our model we only note that at very low temperatures
the electron--phonon coupling becomes weak and this conduction channel can be ignored.
On the other hand, two heat leak mechanisms particular to the type of devices considered
here are heat conduction by electrons moving between the junction and the external circuit,
and photonic heat transfer between the external impedance and the metallic reservoirs of
the junction system. The first leakage channel can be eliminated by using superconducting
wires, and the second one is suppressed if the resistance of the electron reservoirs
is much smaller than the tunnel resistance of the junctions.

We have assumed that the whole electron transport system remains at the same temperature $T$.
This is a nontrivial requirement especially for the small dots which can be easily driven
to different temperatures or even out of equilibrium. However, one central assumption leading
to Eq.~(\ref{currsimp}) was that those junctions which are effectively uncoupled from the environment
have a relatively high transmittance, and the back and forth tunneling of electrons through
these junctions will equilibrate the dot with the reservoir.

One should also note that since the generated current flows through the external
impedance $Z(\omega)$, the power $P$ and voltage bias $V$ must be related
by  $Z(\omega\to0)=V^2/P$ if there are no other voltage sources.
For example, from Fig.~\ref{fig2}(b) we have $Z(0)\sim 10^2 T_0/\Gamma$
which is consistent with the assumption of a high-impedance environment
since the validity of $P(E)$ theory requires $T_0/\Gamma\gg1$.
Of course a non-thermoelectric voltage source in series with the heat engine
can be used to drive the current through an even larger load.

In conclusion, we have investigated the heat engine performance of several types of single-electron
junctions coupled to linear electromagnetic
environments. Highest thermoelectric performance is obtained
when only a single junction in an array of junctions
is coupled to the external heat bath, and in this case a simple formula is able to describe
the essential dynamics. It was confirmed that an energy-selective
junction between two quantum dots is capable of highly efficient energy conversion,
while equally large power production is possible with all studied junction types.

\acknowledgments
One of the authors (T.O.) would like to thank the Academy of Finland for support.

\end{document}